%% file: apssamp.tex
\documentclass[
 reprint,
 superscriptaddress,
 amsmath,amssymb,
 aps,
 prd,
 nofootinbib,
]{revtex4-2}

\usepackage{graphicx}
\usepackage{dcolumn}
\usepackage{bm}
\usepackage{booktabs}
\usepackage{multirow}
\usepackage{array}
\usepackage{xcolor}
\usepackage[
    colorlinks=true,
    linkcolor=blue,
    citecolor=blue,
    urlcolor=black
]{hyperref}

\begin{document}

\preprint{APS/123-QED}

\title{Self-supervised reconstruction of transients in data from space-borne gravitational-wave detectors}

\author{Yuxiang Xu}
\affiliation{Center for Gravitational Wave Experiment, National Microgravity Laboratory, Institute of Mechanics, Chinese Academy of Sciences, Beijing 100190, China}
\affiliation{Taiji Laboratory for Gravitational Wave Universe (Beijing/Hangzhou), University of Chinese Academy of Sciences (UCAS), Beijing 100049, China}
\author{Minghui Du}
\email{duminghui@imech.ac.cn}
\affiliation{Center for Gravitational Wave Experiment, National Microgravity Laboratory, Institute of Mechanics, Chinese Academy of Sciences, Beijing 100190, China}
\author{Bo Liang}
\affiliation{Center for Gravitational Wave Experiment, National Microgravity Laboratory, Institute of Mechanics, Chinese Academy of Sciences, Beijing 100190, China}
\affiliation{Taiji Laboratory for Gravitational Wave Universe (Beijing/Hangzhou), University of Chinese Academy of Sciences (UCAS), Beijing 100049, China}
\author{Zihao Xiao}
\affiliation{Center for Gravitational Wave Experiment, National Microgravity Laboratory, Institute of Mechanics, Chinese Academy of Sciences, Beijing 100190, China}
\affiliation{Taiji Laboratory for Gravitational Wave Universe (Beijing/Hangzhou), University of Chinese Academy of Sciences (UCAS), Beijing 100049, China}
\author{Tianyu Zhao}
\affiliation{Center for Gravitational Wave Experiment, National Microgravity Laboratory, Institute of Mechanics, Chinese Academy of Sciences, Beijing 100190, China}
\author{Peng Xu}
\email{xupeng@imech.ac.cn}
\affiliation{Center for Gravitational Wave Experiment, National Microgravity Laboratory, Institute of Mechanics, Chinese Academy of Sciences, Beijing 100190, China}
\affiliation{Taiji Laboratory for Gravitational Wave Universe (Beijing/Hangzhou), University of Chinese Academy of Sciences (UCAS), Beijing 100049, China}
\affiliation{Hangzhou Institute for Advanced Study, UCAS, Hangzhou 310024, China}
\affiliation{Lanzhou Center of Theoretical Physics, Lanzhou University, Lanzhou 730000, China}

\date{\today}

\begin{abstract}
Space-based gravitational-wave (GW) data may contain transient signals whose waveform morphologies are not known in advance.  Extracting these signals is important for characterizing new sources and mitigating instrumental anomalies.  However, existing deep neural network (DNN)-based extraction approaches rely on clean training targets and waveform-class-specific examples, which may limit their applicability when the transient morphology is not specified in advance.  This work develops a Noise2Noise (N2N)-inspired self-supervised framework that learns from noisy observations without clean training targets and requires no transient-specific waveform templates at inference.  A single model trained on noisy massive black-hole binary (MBHB) observations provides high-overlap MBHB recovery and recovers the dominant morphologies of instrumental glitches and other GW transient signals such as cosmic string bursts in source-confused test data.  Beyond waveform recovery, when independent information identifies a candidate transient as instrumental, its extracted waveform can be subtracted from the data without an anomaly-specific template.  In a simulated continuous data stream, this procedure substantially suppresses the injected-anomaly power within the glitch-dominated frequency band.  These results support the use of self-supervised extraction for initial waveform estimation of candidate transients with unknown morphologies, enabling subsequent characterization and, where appropriate, conditional subtraction of instrumental anomalies.
\end{abstract}

\maketitle

\input{sections/01_introduction.tex}

\input{sections/02_data_preparations.tex}

\input{sections/03_methodology.tex}

\input{sections/04_training_strategy.tex}

\input{sections/05_result.tex}

\input{sections/06_conclusion.tex}

\input{sections/07_acknowledgments.tex}

\bibliographystyle{apsrev4-2}
\bibliography{apssamp}

\end{document}

%% file: sections/01_introduction.tex
\section{Introduction}
\label{sec:introduction}

The first direct observation of gravitational waves, GW150914, by the Advanced LIGO marked the beginning of modern gravitational-wave (GW) astronomy~\cite{abbott2016gw150914}.  The LIGO--Virgo--KAGRA (LVK) Collaboration has since reported hundreds of GW events~\cite{abbott2016observation,abbott2016gw151226,abbott2016binary,scientific2017gw170104,abbott2017gw170608,abbott2017gw170814,abbott2017gw170817,abbott2020gw190425,abbott2020gw190412,abbott2020gw190521,abbott2021tests,ezquiaga2021hearing,bozzola2021general,abbott2019gwtc,abbott2021gwtc,abbott2023gwtc}, substantially advancing our understanding of compact celestial bodies and the Universe.  However, the sensitivities of ground-based GW detectors are intrinsically limited at low frequencies by seismic disturbance, Newtonian noise, and suspension-related disturbances, and are therefore primarily sensitive to signals above approximately $10\,\mathrm{Hz}$~\cite{abbott2019gwtc,abbott2020prospects,freise2010interferometer}.  To explore the low-frequency GW Universe, space-based missions such as the Laser Interferometer Space Antenna (LISA)~\cite{amaro2017laser,baker2019laser}, Taiji~\cite{gong2011scientific,hu2017taiji,liu2026recent}, and TianQin~\cite{luo2016tianqin} are being developed, with designed sensitivities to GW signals in the range from approximately $0.1\,\mathrm{mHz}$ to $1\,\mathrm{Hz}$.  This frequency band is expected to host signals from a diverse set of source classes, including massive black-hole binaries (MBHBs), stellar-origin black-hole binaries (SOBHBs), extreme-mass-ratio inspirals (EMRIs), Galactic white-dwarf binaries, and stochastic gravitational-wave backgrounds (SGWBs).  This low-frequency band offers new observational windows on cosmic evolution, compact-object populations, and new tests of fundamental physics.  It also poses substantial data-analysis challenges because space-based observations involve long-duration signals, overlapping source populations, and complex detector noise.


Whereas cataloged ground-based GW detections to date have been overwhelmingly associated with compact-binary coalescences~\cite{abbott2019gwtc,abbott2021gwtc,abbott2023gwtc}, space-based data are expected to contain many simultaneously present components, including large populations of long-lived Galactic binaries and extreme-mass-ratio inspirals, massive-black-hole mergers, stochastic foregrounds and backgrounds, and potentially burst-like transients~\cite{amaro2017laser,baker2019laser,robson2019detecting,feroz2010classifying}. The coexistence of overlapping resolvable signals and unresolved foregrounds makes purely sequential source-by-source analysis generally inadequate and motivates joint (or ``global'') inference over astrophysical sources and instrumental-noise components~\cite{littenberg2023prototype}. Accurate parameter measurement and physical interpretation in space-based GW astronomy generally rely on long and uninterrupted observations~\cite{baghi2019gravitational,dey2021effect}, which impose stringent requirements on the long-term stability of both the payloads and the spacecraft platform.
During observations of the primary science targets, the data stream may also contain unmodeled burst-like transients~\cite{robson2019detecting,feroz2010classifying}. Such features introduce localized non-Gaussian structure into the data. Instrumental transients can generate spurious triggers~\cite{lee2024impact}, whereas either instrumental or astrophysical transients that are not represented in the analysis model may mask weaker signals or bias inference for overlapping sources~\cite{hourihane2022accurate,antonelli2021noisy}, thereby further complicating global analysis.
Operational experience from LISA Pathfinder (LPF)~\cite{armano2024depth,baghi2022detection,armano2018beyond,armano2022transient,armano2016sub}, the Gravity Recovery and Climate Experiment and its Follow-On mission (GRACE/GFO)~\cite{frommknecht2007integrated,sheard2012intersatellite,flury2008precise,abich2019orbit}, and Taiji-1~\cite{taiji2021china,liu2021orbit,wang2021development,wu2024suppressing} has shown that instrumental transients can affect science data during mission operations.  Representative examples include acceleration glitches associated with the gravitational reference sensor (GRS) system and phase jumps in the interferometer.  Because the physical origins and statistical properties of these disturbances are not yet fully understood~\cite{baghi2022detection,armano2022transient}, they may appear as transient structures in the data stream and interfere with subsequent analyses.  Space-based detectors may also encounter unknown GW sources, either through burst-like signatures, through their contributions to SGWBs~\cite{auclair2020probing,flauger2021improved,auclair2023cosmology}, or through both channels.  Cosmic-string bursts provide a representative example~\cite{auclair2023cosmic,cohen2010searches} and may likewise appear as transient features in space-based GW observations.  The systematic extraction and characterization of these unmodeled transients is therefore of considerable scientific and operational value. Characterizing instrumental transients can aid the monitoring and diagnosis of detector performance, whereas astrophysical or cosmological transients may reveal new source populations or fundamental physics. Accurately modeling and, where appropriate, removing instrumental transient contamination can also reduce non-Gaussian residuals and improve the robustness of downstream analyses, including matched-filter searches and parameter inference.

Representative approaches for detecting, reconstructing, or denoising unmodeled or weakly modeled transients include coherent excess-power and time--frequency clustering methods~\cite{klimenko2016method,drago2021coherent,thrane2013searching,sutton2010x}, Bayesian burst inference~\cite{lynch2017information} and wavelet-based waveform reconstruction~\cite{cornish2015bayeswave,cornish2021bayeswave}, variational denoising with total-variation regularization~\cite{torres2014total,torres2018total}, and sparse-representation methods based on dictionary learning~\cite{torres2016denoising,torres2020application}. Existing unmodeled-transient methods can detect or reconstruct signals without source-specific templates, but rely on explicit signal representations or priors~\cite{klimenko2016method,cornish2015bayeswave,cornish2021bayeswave,torres2016denoising,torres2020application} and remain largely untested in source-confused space-based data~\cite{robson2019detecting,littenberg2023prototype,muratore2025pipeline,houba2025deep}.
A further challenge is that instrumental-noise properties must be inferred in the presence of persistent astrophysical signals, because genuinely source-free ``quiet'' segments may be scarce or unavailable in long-duration space-based observations~\cite{Muratore:2023gxh,csx9-9trp,Aimen:2025zxn}.
Against this background, 
deep neural network (DNN)-based methods have recently been introduced into GW data analysis~\cite{george2018deep,cuoco2021enhancing,zhao2023space,du2024advancing,liang2024rapid,liang2025unlocking,liang2026toward,dax2021real,xu2024gravitational,wang2024waveformer,wei2020gravitational,wang2020gravitational,krastev2021detection,zhang2022detecting,chatterjee2021extraction,bacon2023denoising,murali2023detecting,houba2024detection,dooney2025time}.  For GW transient-signal extraction, DNN-based methods can learn nonlinear mappings from noisy observations to target transient signals through end-to-end training.  These studies have shown the promise of DNNs for GW signal reconstruction and denoising~\cite{zhao2023space,wang2024waveformer,chatterjee2021extraction,bacon2023denoising,dooney2025time}, nonstationary instrumental-noise mitigation~\cite{xu2024gravitational}, and glitch detection or subtraction~\cite{houba2024detection,dooney2025time}.  However, these methods can depend on how well the training data represent observational conditions, limiting their applicability to candidate transients whose waveform morphologies are not specified in advance~\cite{cuoco2021enhancing,schafer2022training}.

To address this challenge, we develop a Noise2Noise (N2N)~\cite{lehtinen2018noise2noise}-inspired self-supervised framework for transient-waveform extraction from space-based GW data.  The method learns directly from noisy observations, requiring neither clean training targets nor transient-specific waveform templates at inference.  It combines a time-domain encoder--decoder with a bidirectional recurrent bottleneck and a random neighboring-subsample consistency objective.  We train a single model using noisy MBHB observations and evaluate it on MBHBs, instrumental glitches, and cosmic-string bursts in simulated source-confused data as typical transients.  The latter two classes test cross-morphology waveform extraction when the target morphology is not specified during training.  Finally, for anomalies independently identified as instrumental, we evaluate conditional subtraction in a simulated 30-day stream.  Overall, the framework provides initial waveform estimates for candidate transients with unknown morphologies, supporting subsequent source characterization and, where appropriate, conditional subtraction of instrumental anomalies.

%% file: sections/02_data_preparations.tex
\section{Data preparation}
\label{sec:data_preparations}
For space-based GW detection missions such as LISA, Taiji, and TianQin, the raw inter-spacecraft measurements are processed using the time-delay interferometry (TDI) technique to suppress the dominant laser frequency noise~\cite{tinto2021time,armstrong1999time,babak2021lisa}, and the scientific analysis of GW signals  begins with the  resulting TDI data streams.
To achieve effective  laser noise suppression and optimal  GW signal sensitivity, 
throughout this work, we use the Michelson-$A_2$ channel of the TDI observables~\cite{PhysRevD.66.122002,Tinto:2022zmf}. 
Each simulated observation can be expressed  as
\begin{equation}
x(t)=s(t)+n(t),
\end{equation}
where $s(t)$ denotes the TDI response  to  GW  signal (for MBHBs, cosmic-string bursts, etc.) or instrumental transient (for glitches),  and $n(t)$ denotes the  noise in the TDI channel. 
After the suppression of laser frequency noise via TDI, the residual instrumental noise budget is dominated by test-mass acceleration (ACC) noise from the GRS and optical metrology system (OMS) noise. 
In this paper, we take Taiji as an example for the space-based GW detectors.  Based on the baseline design of Taiji~\cite{luo2020brief}, we adopt the amplitude spectral density (ASD) models: 
\begin{equation}
S_{\rm OMS}^{1/2}(f)=8\times10^{-12}\left[1+\left(\frac{2\,\mathrm{mHz}}{f}\right)^4\right]^{1/2}\,\mathrm{m}\,\mathrm{Hz}^{-1/2},
\label{eq:soms}
\end{equation}
\begin{equation}
\begin{split}
S_{\rm ACC}^{1/2}(f)=&\,3\times10^{-15}
\left[1+\left(\frac{0.4\,\mathrm{mHz}}{f}\right)^2\right]^{1/2}\\
&\times
\left[1+\left(\frac{f}{8\,\mathrm{mHz}}\right)^4\right]^{1/2}
\mathrm{m}\,\mathrm{s}^{-2}\,\mathrm{Hz}^{-1/2}.
\end{split}
\label{eq:sacc}
\end{equation}
which are used for instrumental noise generation,  signal-to-noise ratio (SNR) normalization and data whitening.

Data simulation is performed using the  Taiji Data Challenge (TDC) toolkit \texttt{Triangle}~\cite{du2026towards}.
For the detector orbit, we adopt the numerically  simulated  unequal-arm orbit of TDC. 
For instrumental noise, both ACC and OMS noises are generated as Gaussian stochastic processes following  Eq.~(\ref{eq:sacc}) and Eq.~(\ref{eq:soms}) and combined into the TDI data stream via the  \texttt{Triangle-Simulator}~\footnote{https://github.com/TriangleDataCenter/Triangle-Simulator} module. 
MBHB signals are injected using the  \texttt{Triangle-BBH}~\footnote{https://github.com/TriangleDataCenter/Triangle-BBH} module,  which employs the \texttt{IMRPhenom} waveform implemented in  \texttt{BBHx}~\cite{katz2020gpu} and \texttt{WF4PY}~\cite{iacovelli2022gwfast}. 
The waveform is propagated through the numerical response of the space-based detector,  converted into TDI observables, and then   added to the instrumental noise.  
The parameter space used for the MBHB dataset is summarized in Table~\ref{tab:mbhb_parameters}.

\begin{table}[!htbp]
\caption{Parameter ranges used for MBHB signal generation.}
\label{tab:mbhb_parameters}
\begin{ruledtabular}
\renewcommand{\arraystretch}{1.2}
\begin{tabular}{lcc}
Parameter & Lower bound & Upper bound \\
\colrule
$M_{\rm c}$\footnote{Chirp mass of the binary system.} & $10^5 M_{\odot}$ & $10^7 M_{\odot}$ \\
$q$\footnote{Mass ratio $q=m_2/m_1\leq1$.} & 0.01 & 1 \\
$s_1^z$\footnote{Aligned dimensionless spin of the primary black hole.} & $-0.99$ & 0.99 \\
$s_2^z$\footnote{Aligned dimensionless spin of the secondary black hole.} & $-0.99$ & 0.99 \\
$\iota$\footnote{Inclination angle of the orbital plane with respect to the line of sight.} & 0 & $\pi$ \\
$\beta$\footnote{Ecliptic latitude of the source.} & $-\pi/2$ & $\pi/2$ \\
$\lambda$\footnote{Ecliptic longitude of the source.} & 0 & $2\pi$ \\
$\psi$\footnote{Polarization angle of the GW signal.} & 0 & $\pi$ \\
$\phi_{\rm c}$\footnote{Coalescence phase in radians.} & 0 & $2\pi$ \\
$t_{\rm c}$\footnote{Coalescence time within the one-year mission interval.} & $0.01\,\mathrm{yr}$ & $1.0\,\mathrm{yr}$ \\
$d_{\rm L}$\footnote{Luminosity distance of the source.} & $6\times10^3\,\mathrm{Mpc}$ & $10^5\,\mathrm{Mpc}$ \\
\end{tabular}
\end{ruledtabular}
\end{table}

Instrumental glitches are also simulated via \texttt{Triangle-Simulator}, for which  we adopt the  LPF-inspired model for GRS glitches~\cite{armano2022transient}.  The acceleration transient is modeled as
\begin{equation}
g(t)=\frac{\Delta v}{\tau_1-\tau_2}
\left[
\exp\left(-\frac{t-t_0}{\tau_1}\right)-
\exp\left(-\frac{t-t_0}{\tau_2}\right)
\right]\Theta(t-t_0),
\label{eq:lpf}
\end{equation}
where $t$ is the time coordinate of the simulated segment, $t_0$ is the glitch-injection time, $\Delta v$ is the impulse transferred by the glitch, $\tau_1$ and $\tau_2$ are characteristic time scales, and $\Theta(t-t_0)$ denotes the Heaviside step function.  The simulated glitch is injected into the GRS acceleration-noise channel and then propagated through the interferometric measurements and TDI response.  The parameter ranges used for glitch generation are summarized in Table~\ref{tab:glitch_parameters}.

\begin{table}[!htbp]
\caption{Parameter ranges used for LPF-like GRS glitch generation.}
\label{tab:glitch_parameters}
\begin{ruledtabular}
\renewcommand{\arraystretch}{1.2}
\begin{tabular}{lcc}
Parameter & Lower bound & Upper bound \\
\colrule
$\Delta v$\footnote{Velocity-kick amplitude of the acceleration glitch.} & $1.18\times10^{-12}\,\mathrm{m}\,\mathrm{s}^{-1}$ & $2.2\times10^{-12}\,\mathrm{m}\,\mathrm{s}^{-1}$ \\
$\tau_1$\footnote{First characteristic time scale of the glitch.} & $10.0\,\mathrm{s}$ & $5661.65\,\mathrm{s}$ \\
$\tau_2$\footnote{Second characteristic time scale of the glitch.} & $\tau_1$ & $5661.71\,\mathrm{s}$ \\
$t_0$\footnote{Glitch-injection time within the one-year mission interval.} & $0.01\,\mathrm{yr}$ & $1.0\,\mathrm{yr}$ \\
\end{tabular}
\end{ruledtabular}
\end{table}

For the simulation of cosmic-string bursts, we follow the cusp-burst waveform prescription of Ref.~\cite{auclair2023cosmic}.   The two GW polarizations are then given by
\begin{equation}
h_+(t)=\cos(2\psi)h(t),\qquad
h_\times(t)=\sin(2\psi)h(t),
\label{eq:cosmic_string_polarizations}
\end{equation}
where $\psi$ is the polarization angle and the function $h(t)$ is given by
\begin{equation}
h(t)=2A\int_{f_{\rm low}}^{f_{\rm high}} f^{-4/3}
\cos\left[2\pi f\left(t-t_{\rm c}^{\rm cs}\right)\right]df,
\label{eq:cosmic_string_waveform}
\end{equation}
where $A$ is the waveform amplitude, $f_{\rm low}$ and $f_{\rm high}$ define the frequency band of the burst, and $t_{\rm c}^{\rm cs}$ denotes the cusp peak time. 
The TDI responses of the signals are computed with \texttt{Triangle-Simulator}. 
The parameter ranges used for the cosmic-string burst simulations are summarized in Table~\ref{tab:cosmic_string_parameters}.

\begin{table}[!htbp]
\caption{Parameter ranges used for cosmic-string burst generation.}
\label{tab:cosmic_string_parameters}
\begin{ruledtabular}
\renewcommand{\arraystretch}{1.2}
\begin{tabular}{lcc}
Parameter & Lower bound & Upper bound \\
\colrule
$A$\footnote{Waveform amplitude parameter, in $\mathrm{Hz}^{1/3}$.} & $10^{-22}$ & $10^{-19}$ \\
$f_{\rm low}$\footnote{Lower frequency cutoff of the cusp burst.} & $10^{-5}\,\mathrm{Hz}$ & $10^{-4}\,\mathrm{Hz}$ \\
$f_{\rm high}$\footnote{Upper frequency cutoff of the cusp burst.} & $10^{-3}\,\mathrm{Hz}$ & $10^{-2}\,\mathrm{Hz}$ \\
$\lambda$\footnote{Ecliptic longitude of the source.} & 0 & $2\pi$ \\
$\beta$\footnote{Ecliptic latitude of the source, sampled isotropically by drawing $\sin\beta$ uniformly.} & $-\pi/2$ & $\pi/2$ \\
$\psi$\footnote{Polarization angle of the GW signal.} & 0 & $\pi$ \\
$t_{\rm c}^{\rm cs}$\footnote{Cusp peak time within one year.} & $0.01\,\mathrm{yr}$ & $1.0\,\mathrm{yr}$ \\
\end{tabular}
\end{ruledtabular}
\end{table}

For evaluation only, we directly use the GB foreground and SGWB realizations from the TDC II data release~\cite{du2026towards}, and process them to match the $A_2$-channel convention and sampling frequency used for the other simulated components. These components are entirely absent from the training set and are treated as astrophysical foreground and background contamination rather than instrumental noise or extraction targets. Their inclusion is intended to evaluate the robustness of the model in the presence of persistent source confusion and components outside the training distribution.

Each training sample consists of a one-day data segment sampled at $0.1\,\mathrm{Hz}$, corresponding to 8640 time-domain samples per segment.  Each target signal or transient response is rescaled to a prescribed optimal SNR before being added to the noise.  For a target $s$, the optimal SNR is defined as
\begin{equation}
\rho=(s|s)^{1/2},
\label{eq:optimal_snr}
\end{equation}
where the noise-weighted inner product is
\begin{equation}
(a|b)=2\int_{f_{\rm min}}^{f_{\rm max}}
\frac{\tilde a(f)\tilde b^{*}(f)+\tilde a^{*}(f)\tilde b(f)}
{S_{A_2}(f)}\,df.
\label{eq:noise_weighted_inner_product}
\end{equation}
Here, $S_{A_2}(f)$ is the noise PSD of the $A_2$ TDI channel, the tilde denotes the Fourier transform, and the superscript $*$ denotes complex conjugation.  We set $f_{\rm min}=10^{-5}\,\mathrm{Hz}$ and $f_{\rm max}=0.05\,\mathrm{Hz}$, corresponding to the Nyquist frequency of the sampled data.  The same inner product is used to quantify extraction performance.  For an extracted output $o$ and the corresponding reference waveform $h$, the overlap is defined as
\begin{equation}
\mathcal{O}(o,h)=
\max_{t_{\rm c},\phi_{\rm c}}
\left(\hat{o}\middle|\hat{h}\right),
\label{eq:overlap}
\end{equation}
where $\hat{o}$ and $\hat{h}$ are the normalized versions of $o$ and $h$, respectively, and $t_{\rm c}$ and $\phi_{\rm c}$ denote the time and phase shifts at which the overlap between $o$ and $h$ is maximized. 

The training set contains only MBHB signals injected into instrumental noise.
The predicted SNRs of MBHBs  based on astrophysical population models can be up to $\mathcal{O}(10^3)$~\cite{2017arXiv170200786A},  
while for these brightest MBHBs the detection and identification would be trivial tasks, therefore in this study the  target SNRs of our training waveforms are  sampled in the range $30\leq\rho\leq50$.

%% file: sections/03_methodology.tex
\section{Methodology}
\label{sec:methodology}

\begin{figure*}[t]
\centering
\includegraphics[width=\textwidth]{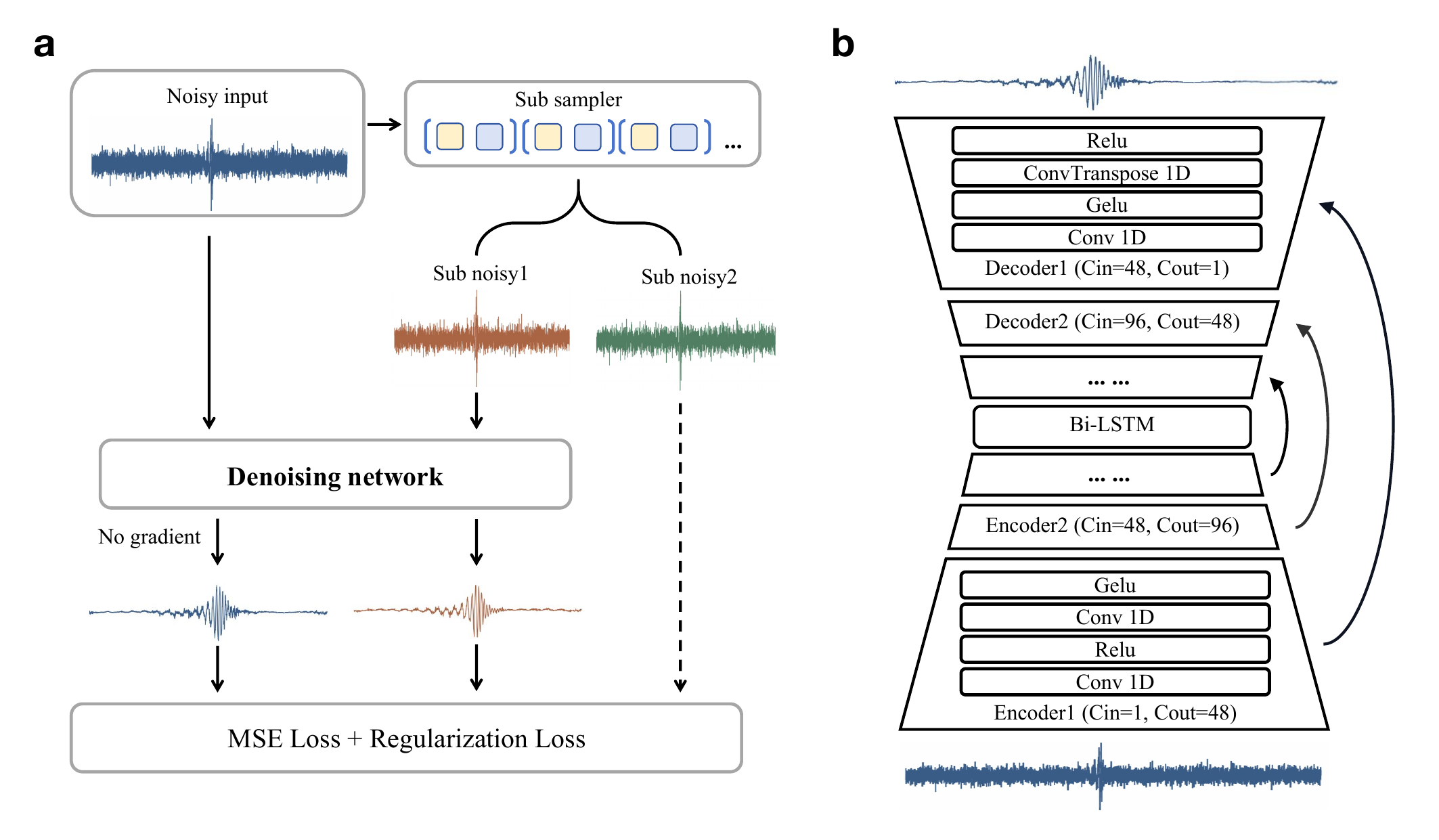}
\caption{Overview of the proposed self-supervised transient-signal extraction framework.  Panel (a) shows the overall pipeline, including the construction of neighboring subsamples from a whitened noisy observation, self-supervised optimization, and extraction of transient signals.  Panel (b) shows the detailed architecture of the time-domain extractor, which consists of convolutional encoding and decoding blocks, skip connections, and a bidirectional recurrent bottleneck.}
\label{fig:method_overview}
\end{figure*}

The proposed framework is summarized in Fig.~\ref{fig:method_overview}.  We first review the basic principle of N2N denoising as the methodological starting point, and then describe the neighboring-subsample objective and time-domain network used to implement transient-signal extraction in space-based GW data.

\subsection{Principle of N2N denoising}
\label{subsec:n2n_principle}

The central idea of N2N denoising is that clean targets are not strictly required when two independent noisy observations of the same underlying signal are available.  Let
\begin{equation}
y=s+n,\qquad y'=s+n',
\label{eq:n2n_pair}
\end{equation}
where $s$ is the underlying signal and $n$ and $n'$ are independent zero-mean noise realizations.  A denoising network can be trained by minimizing
\begin{equation}
\mathcal{L}_{\rm N2N}(\theta)
=\mathbb{E}_{y,y'}\left[\left\|f_\theta(y)-y'\right\|_2^2\right].
\label{eq:n2n_loss}
\end{equation}
Because the noise in the target $y'$ is independent of the input and has zero mean, the target noise is averaged out in expectation.  Consequently, the optimal estimator under the mean-squared-error objective is
\begin{equation}
f_\theta^\ast(y)=\mathbb{E}[y'|y]=\mathbb{E}[s|y],
\label{eq:n2n_expectation}
\end{equation}
which is the same conditional mean estimated in supervised denoising with clean targets.

\subsection{Random neighboring-subsample objective}
\label{subsec:self_supervised_objective}

The standard N2N formulation relies on paired measurements of the same signal with statistically independent noise.  Such pairs are generally unavailable in space-based GW data, where transients occur in a single continuous and source-confused stream.
In the simulated training data, the instrumental noise is Gaussian by construction, and whitening with the corresponding instrumental-noise PSD reduces its temporal correlations. Meanwhile, the MBHB training signals generally evolve smoothly relative to the adopted sampling interval of $10\,\mathrm{s}$, so neighboring samples tend to retain similar local signal structure. Therefore, when the residual correlation between neighboring whitened-noise samples and the local variation of the clean signal are both sufficiently small, complementary neighboring views provide an approximate realization of the shared-signal and independent-noise conditions underlying N2N.
We therefore approximate the paired-view requirement by randomly splitting neighboring samples of one observation into two complementary half-rate sequences.  This construction preserves local signal coherence while avoiding an identical input--target pair.

Let $x\in\mathbb{R}^{T}$ denote a whitened noisy input sequence with even length $T$.  The proposed objective constructs two complementary half-rate views by randomly splitting each adjacent pair of samples.  For the $k$th adjacent pair, let $b_k\sim\mathrm{Bernoulli}(1/2)$.  The two views are defined as
\begin{equation}
\begin{split}
v_{1,k} &= b_k x_{2k}+(1-b_k)x_{2k+1},\\
v_{2,k} &= (1-b_k)x_{2k}+b_k x_{2k+1},
\end{split}
\label{eq:neighbor_subsampling}
\end{equation}
for $k=0,\ldots,T/2-1$.  Thus, $v_1$ and $v_2$ contain complementary samples from the same local waveform neighborhood.  This construction discourages the network from simply copying the input sample by sample while preserving the local coherence of transient signals.

The first loss term asks the extractor to predict one neighboring view from the other,
\begin{equation}
\mathcal{L}_{\rm pred}
=\frac{2}{T}\left\|f_\theta(v_1)-v_2\right\|_2^2.
\label{eq:prediction_loss}
\end{equation}
To make the half-rate training task consistent with the full-resolution extraction operator used at inference time, we also apply the model to the full input sequence without propagating gradients through this branch.  Let
\begin{equation}
\bar{x}=\operatorname{stopgrad}\left[f_\theta(x)\right],
\end{equation}
and let $\bar{v}_1$ and $\bar{v}_2$ be two complementary neighboring subsamples of $\bar{x}$ constructed with the same random-subsampling rule as in Eq.~\eqref{eq:neighbor_subsampling}.  The consistency regularization is then
\begin{equation}
\mathcal{L}_{\rm reg}
=\frac{2}{T}\left\|f_\theta(v_1)-v_2-\bar{v}_1+\bar{v}_2\right\|_2^2.
\label{eq:regularization_loss}
\end{equation}
This term constrains the residual learned in the neighboring-view prediction task, $f_\theta(v_1)-v_2$, to be consistent with the residual implied by the full-sequence extracted estimate, $\bar{v}_1-\bar{v}_2$.  This term is intended to suppress view-specific noise fitting in the half-rate training task and to align the self-supervised objective with the full-resolution extraction operator used at inference time.

The total training objective is
\begin{equation}
\mathcal{L}=\mathcal{L}_{\rm pred}+\gamma\mathcal{L}_{\rm reg},
\label{eq:self_supervised_loss}
\end{equation}
where we set $\gamma=1$ in this work.  The first term provides the N2N-style neighboring-view prediction signal, whereas the second term regularizes the prediction by tying it to the full-sequence extracted output.  Together, these two terms define the self-supervised objective used to optimize the extraction network.

\subsection{Time-domain extraction network}
\label{subsec:network}

We adopt a time-domain autoencoder architecture for the signal-extraction operator $f_\theta$.  The network takes a whitened one-dimensional sequence as input.  This sequence is further normalized by its sample standard deviation before being processed by the network and is rescaled back to the original amplitude after reconstruction.  For notational simplicity, $x\in\mathbb{R}^{T}$ denotes the normalized input below.

The network is built from downsampling blocks, upsampling blocks, and a recurrent bottleneck.  A downsampling block first applies a strided one-dimensional convolution to reduce the temporal resolution and increase the feature dimension, then uses a rectified linear unit (ReLU), a pointwise convolution, and a gated linear unit (GLU) to refine the local representation.  We write the $\ell$th downsampling block as
\begin{equation}
E_\ell(y)=\operatorname{GLU}\left[
C_{\ell}^{1}\left(
\operatorname{ReLU}\left(C_{\ell}^{k,s}(y)\right)
\right)\right],
\label{eq:encoder_block}
\end{equation}
where $C_{\ell}^{k,s}$ denotes a one-dimensional convolution with kernel size $k=8$ and stride $s=4$, and $C_{\ell}^{1}$ denotes a pointwise convolution.  The initial channel width is set to 48 and is doubled after each downsampling stage.

The upsampling block mirrors the downsampling block.  It first applies a pointwise gated convolution to mix channel-wise information and then uses a transposed convolution to restore the temporal resolution.  The corresponding block is expressed as
\begin{equation}
D_\ell(y)=T_{\ell}^{k,s}\left[
\operatorname{GLU}\left(\tilde{C}_{\ell}^{1}(y)\right)
\right],
\label{eq:decoder_block}
\end{equation}
where $\tilde{C}_{\ell}^{1}$ is a pointwise convolution and $T_{\ell}^{k,s}$ is a transposed convolution with the same kernel size and stride as the associated downsampling block.  Between the encoder and decoder, the compressed feature sequence is processed by a two-layer bidirectional long short-term memory (BLSTM) bottleneck,
\begin{equation}
B(y)=P\left[\operatorname{BLSTM}_{2}(y)\right],
\label{eq:bottleneck_block}
\end{equation}
where $P$ denotes the linear projection used to map the bidirectional recurrent features back to the decoder feature dimension.  This bottleneck introduces temporal context after convolutional compression, while the convolutional blocks preserve local waveform morphology.

Using these building blocks, the complete network is constructed in an encoder--bottleneck--decoder form.  The normalized input sequence is first upsampled by a factor of four with sinc-interpolation filters and is then passed through five downsampling blocks to obtain a compressed temporal representation.  This representation is processed by the BLSTM bottleneck to incorporate temporal context beyond the local receptive field of the convolutional layers.  The decoder then applies the corresponding upsampling blocks in reverse order to produce the extracted time series.  At each decoding stage, the feature map is combined with the encoder feature map at the same resolution through a skip connection, with boundary cropping applied when necessary.  These skip connections reduce information loss during compression and help preserve the local morphology of transient signals.  Finally, the reconstructed sequence is downsampled by the corresponding sinc filters, cropped to the original length, and rescaled to the original amplitude.

%% file: sections/04_training_strategy.tex
\section{Training strategy}
\label{sec:training_strategy}

The model parameters are optimized with the Adam optimizer~\cite{kingma2014adam}.  We adopt a warmup phase~\cite{he2016deep} followed by a cosine-annealing learning-rate schedule~\cite{loshchilov2016sgdr} to stabilize early training.  The maximum learning rate is set to $3\times10^{-4}$.  During the first 1000 optimization steps, the learning rate is linearly increased from zero to this maximum value.  After the warmup stage, the learning rate is gradually decayed according to a cosine-annealing schedule.  The network is trained for 100 epochs with a batch size of 64.  The overall training framework is implemented in PyTorch~\cite{paszke2019pytorch}, and the training is performed on four NVIDIA A100 40~GB graphics processing units (GPUs) for approximately 6 hours.

%% file: sections/05_result.tex
\section{Results}
\label{sec:result}

We evaluated the proposed method on three transient classes: MBHB signals, instrumental glitches, and cosmic-string bursts.  The analysis covered MBHB recovery in source-confused data and cross-morphology extraction of the latter two classes, whose morphologies were not specified to the model in advance.  We also evaluated conditional subtraction of candidate transients independently identified as instrumental.

\begin{figure}[!t]
\centering
\includegraphics[width=0.96\columnwidth]{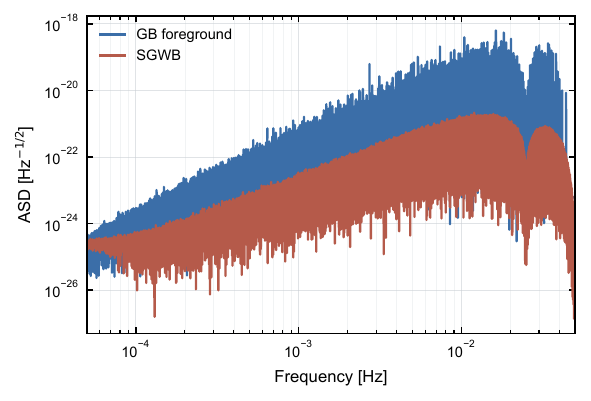}
\caption{ASDs of the additional unresolved foreground/background components used in the evaluation data.  These components are added to the instrumental-noise realizations in the test sets to emulate a more source-confused space-based GW observation environment.}
\label{fig:background_asd}
\end{figure}

For each class, we generated 1,000 samples at each of three optimal SNRs: 30, 40, and 50.  For MBHBs, these values probe the relatively low-SNR regime expected for space-based observations, providing a challenging test of weak-signal recovery.  We used the same SNRs for glitches and cosmic-string bursts to control signal strength and enable direct comparisons across waveform classes.  To emulate source-confused observations, all test sets also included the GB foreground and SGWB components from TDC II~\cite{du_2025_15532090}\footnote{\url{https://doi.org/10.5281/zenodo.15532090}}.  Figure~\ref{fig:background_asd} shows the ASDs of these two additional background components in the $A_2$ channel.

\subsection{MBHB waveform recovery}
\label{subsec:mbhb_waveform_recovery}

\begin{figure*}[!t]
\centering
\includegraphics[width=\textwidth]{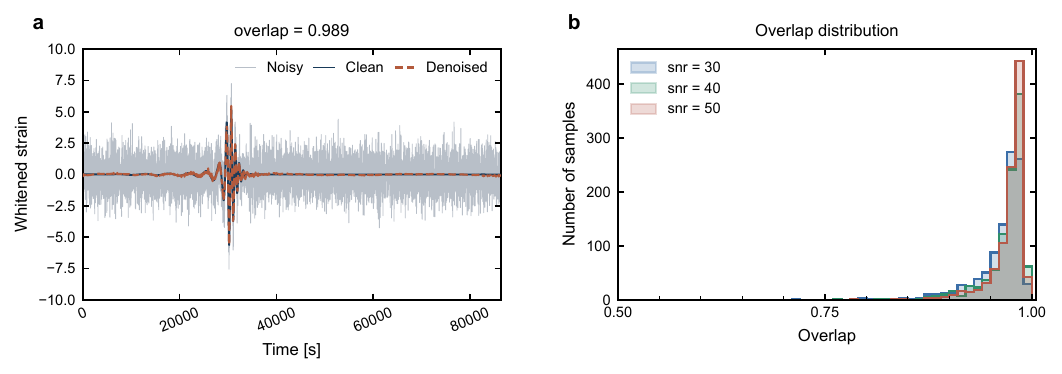}
\caption{MBHB waveform-recovery performance.  Panel (a) shows a representative extraction example, comparing the noisy input, the template waveform, and the extracted output.  Panel (b) shows the overlap distributions between the extracted outputs and the corresponding template waveforms for MBHB test sets with SNRs of 30, 40, and 50.}
\label{fig:mbhb_extraction}
\end{figure*}

We first evaluate MBHB waveform recovery in test segments that include the additional GB and SGWB backgrounds described above.  The results are shown in Fig.~\ref{fig:mbhb_extraction}.  Figure~\ref{fig:mbhb_extraction}(a) presents a randomly selected extraction example.  The extracted waveform closely follows the corresponding template waveform, indicating that both the phase evolution and amplitude modulation of the MBHB signal are well recovered.

We quantified recovery across the full MBHB test sets using the overlap $\mathcal{O}$ between each extracted waveform and its corresponding template.  Because MBHB signals were represented during training, we adopted $\mathcal{O}=0.9$ as a stringent empirical reference level.  As shown in Fig.~\ref{fig:mbhb_extraction}(b), 94.8\%, 97.3\%, and 98.4\% of samples exceeded this level at SNRs of 30, 40, and 50, respectively.  The fraction of high-overlap recoveries therefore increased with SNR, indicating progressively more consistent MBHB extraction in the simulated source-confused data.

\subsection{Cross-morphology waveform extraction}
\label{subsec:cross_morphology_extraction}

We next evaluated cross-morphology waveform extraction for glitches and cosmic-string bursts, neither of which was represented during training.  As in the MBHB tests, the test sets included the additional GB foreground and SGWB components.  Because cross-morphology recovery presents a more demanding generalization task, we adopted a lower empirical reference level of $\mathcal{O}=0.8$ for both classes.

\begin{figure*}[!t]
\centering
\includegraphics[width=\textwidth]{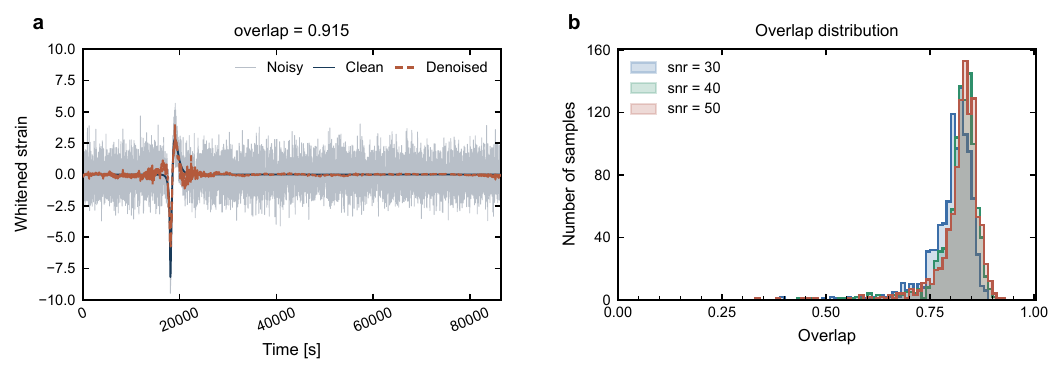}
\caption{Cross-morphology waveform extraction for instrumental glitches.  Panel (a) shows a representative extraction example, comparing the noisy input, the injected reference waveform, and the extracted output.  Panel (b) shows the overlap distributions between the extracted outputs and the corresponding reference waveforms for glitch test sets with SNRs of 30, 40, and 50.}
\label{fig:glitch_extraction}
\end{figure*}

The glitch results are shown in Fig.~\ref{fig:glitch_extraction}.  The extracted waveform in Fig.~\ref{fig:glitch_extraction}(a) retained an identifiable glitch morphology, although recovery was moderately degraded relative to the MBHB reference case.  Across the full glitch test sets, Fig.~\ref{fig:glitch_extraction}(b) shows that 66.2\%, 79.6\%, and 81.1\% of samples exceeded $\mathcal{O}=0.8$ at SNRs of 30, 40, and 50, respectively.  These fractions increased with SNR, indicating improved glitch recovery at higher signal strengths.

\begin{figure*}[!t]
\centering
\includegraphics[width=\textwidth]{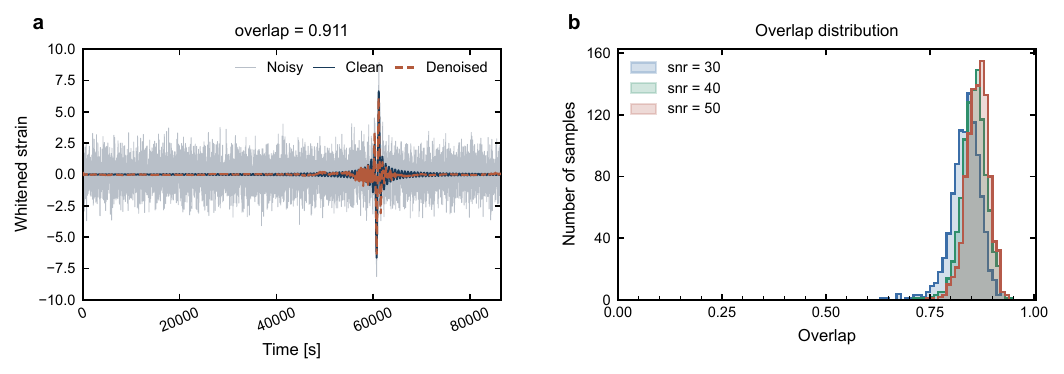}
\caption{Cross-morphology waveform extraction for cosmic-string bursts.  Panel (a) shows a representative extraction example, comparing the noisy input, the injected reference waveform, and the extracted output.  Panel (b) shows the overlap distributions between the extracted outputs and the corresponding reference waveforms for cosmic-string burst test sets with SNRs of 30, 40, and 50.}
\label{fig:string_extraction}
\end{figure*}

The cosmic-string burst results are shown in Fig.~\ref{fig:string_extraction}.  The example in Fig.~\ref{fig:string_extraction}(a) shows that the model recovered the dominant burst structure.  Across the full cosmic-string burst test sets, Fig.~\ref{fig:string_extraction}(b) shows that 87.1\%, 96.8\%, and 98.3\% of samples exceeded $\mathcal{O}=0.8$ at SNRs of 30, 40, and 50, respectively.

The higher overlap fractions for cosmic-string bursts may partly reflect the comparatively regular cusp-like structure of the simulated bursts.  In contrast, the glitches span a broader range of rise and decay timescales.  Across both cases, the same self-supervised extractor recovered the dominant transient structures without clean targets during training or transient-specific waveform templates at inference.

\subsection{Neighboring-subsample consistency analysis}
\label{subsec:subsample_consistency}

\begin{figure*}[!t]
\centering
\includegraphics[width=\textwidth]{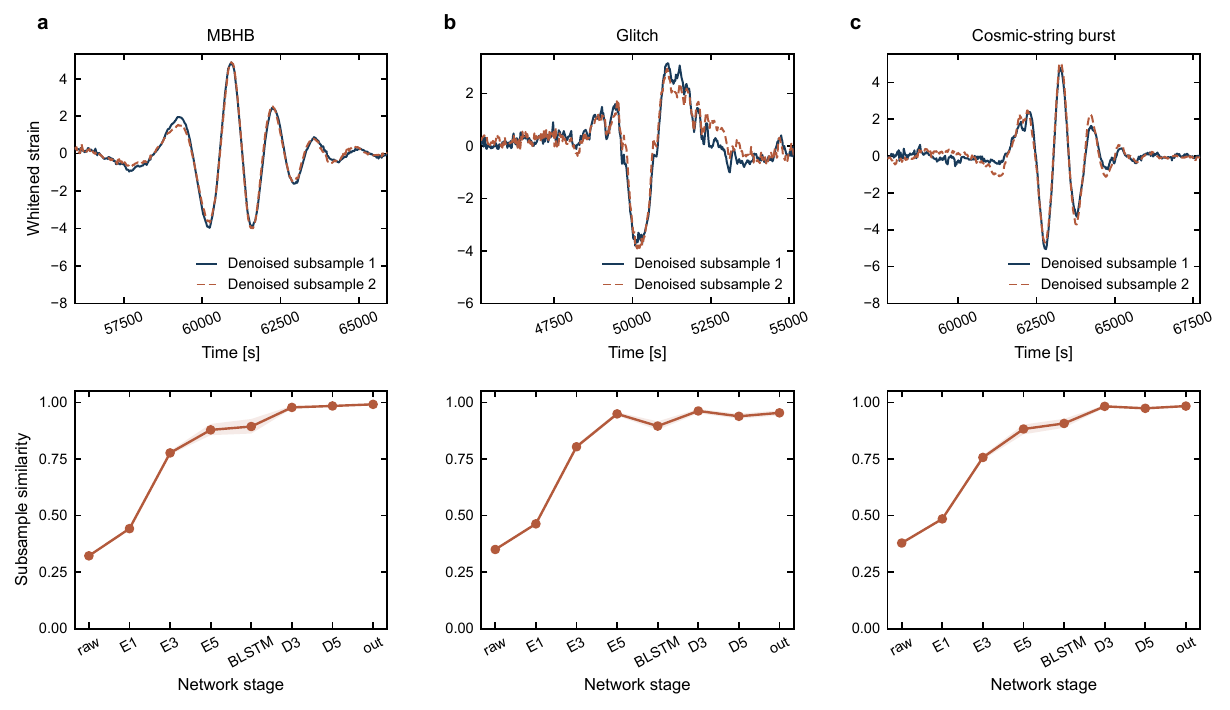}
\caption{Neighboring-subsample consistency analysis.  For each signal class, the upper panel compares the extracted outputs obtained from two complementary neighboring subsamples of the same noisy observation.  The lower panel shows the cosine similarity between the two subsample representations across different network stages.  The increasing similarity from the raw input to the final output indicates that the network progressively maps complementary noisy views toward a common signal-coherent representation.}
\label{fig:subsample_consistency}
\end{figure*}

To gain insight into the behavior of the self-supervised objective, we examine whether complementary neighboring subsamples are mapped to consistent signal representations.  For each noisy observation, two half-rate views are constructed using the random neighboring-subsample rule in Sec.~\ref{subsec:self_supervised_objective}.  These views differ in sample-level fluctuations but share the same local coherent transient structure.  As shown in the upper panels of Fig.~\ref{fig:subsample_consistency}, the extracted outputs from the two views agree well in the main transient features for MBHB and cosmic-string burst examples, while the glitch example shows a moderate but still coherent reconstruction of the principal temporal structure.

We quantify this behavior by computing the cosine similarity between the two subsample representations at different network stages,
\begin{equation}
\mathcal{S}_{\ell}
=
\frac{
\left\langle \phi_{\ell}(v_1),\phi_{\ell}(v_2)\right\rangle
}{
\left\|\phi_{\ell}(v_1)\right\|
\left\|\phi_{\ell}(v_2)\right\|
},
\label{eq:subsample_similarity}
\end{equation}
where $\phi_{\ell}(\cdot)$ denotes the representation at the $\ell$th network stage.  The lower panels of Fig.~\ref{fig:subsample_consistency} show that the similarity is relatively low at the raw-input level and increases through the encoder, the BLSTM bottleneck, and the decoder, reaching high values at the final output.  This behavior is consistent with the network suppressing view-specific noise fluctuations and mapping complementary noisy views toward a common signal-coherent representation, thereby supporting the neighboring-subsample training strategy.

\subsection{Conditional instrumental-anomaly subtraction}
\label{subsec:continuous_stream_subtraction}

To assess the applicability of the proposed method to long-duration data, we construct a 30-day simulated observation stream containing instrumental noise, the GB foreground, and the SGWB component.  
We then inject 30 LPF-like GRS glitches as representative instrumental anomalies.
Previous analyses of LISA Pathfinder data showed that instrumental glitches can be modeled and subtracted to reduce transient contamination~\cite{armano2022transient,baghi2022detection}.  Here, we consider a complementary setting in which auxiliary data or payload-monitoring data have independently identified a candidate transient as instrumental, although its waveform remains unknown.  Assuming that the anomaly contains no astrophysical or cosmological signal, we test whether the proposed framework can estimate and subtract it without an anomaly-specific waveform template.
The contaminated sequence is divided into one-day segments and processed by the trained extraction network.  
The extracted anomaly estimates are then subtracted from the contaminated stream to obtain an anomaly-subtracted time series.

Because the anomaly-free baseline is available in this simulation, the injected anomaly $g(t)$ and the post-subtraction residual $r(t)$ can be evaluated directly.  Here, $r(t)=g(t)-\hat{g}(t)$, where $\hat{g}(t)$ is the anomaly estimate produced by the network.  For a frequency band $\mathcal{B}$, we define
\begin{align*}
P_x(\mathcal{B}) &= \int_{\mathcal{B}}\mathcal{A}_x^2(f)\,df, \\
\eta_{\mathrm{sup}}(\mathcal{B}) &= 1-\frac{P_r(\mathcal{B})}{P_g(\mathcal{B})},
\end{align*}
where $\mathcal{A}_x(f)$ is the ASD of $x(t)$ and $\eta_{\mathrm{sup}}$ is the fractional suppression of the injected-anomaly power.  All components are evaluated using the same spectral-estimation settings, and their power levels relative to the baseline are reported through $P_x(\mathcal{B})/P_0(\mathcal{B})$.

\begin{figure*}[!t]
\centering
\includegraphics[width=\textwidth]{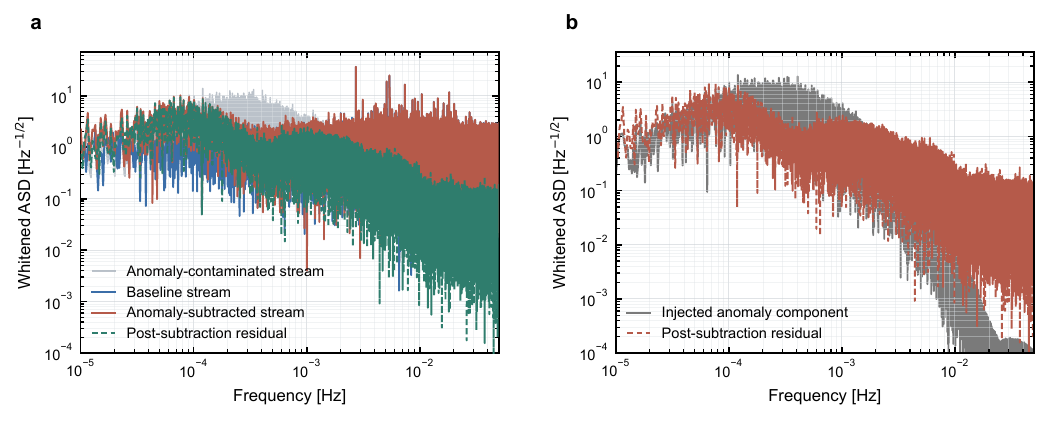}
\caption{Conditional subtraction of instrumental anomalies in a continuous 30-day simulated observation stream.  Panel (a) compares the ASDs of the baseline stream, the anomaly-contaminated stream, the anomaly-subtracted stream, and the post-subtraction residual.  Panel (b) compares the ASD of the injected anomaly component with the post-subtraction residual.}
\label{fig:continuous_stream_subtraction}
\end{figure*}

As shown in Fig.~\ref{fig:continuous_stream_subtraction}, subtraction moves the contaminated stream toward the anomaly-free baseline and reduces the residual relative to the injected anomalies.  Over the full analyzed frequency range of $10^{-5}$--$5\times10^{-2}\,\mathrm{Hz}$, the injected-anomaly power is suppressed by $69.8\%$, and the residual power is $7.7\%$ of the baseline power.  Within the glitch-dominated band of $10^{-4}$--$10^{-3}\,\mathrm{Hz}$, the suppression reaches $87.1\%$, leaving $12.9\%$ of the injected-anomaly power in the residual.  In this band, subtraction reduces the contaminated-stream power from $14.13$ to $2.71$ times the baseline power.

%% file: sections/06_conclusion.tex
\section{Conclusion}
\label{sec:conclusion}

We developed an N2N-inspired self-supervised framework for extracting transient waveforms from space-based GW data.  The framework learns directly from noisy observations without clean training targets and requires no transient-specific waveform templates at inference.  It is therefore suited to initial waveform estimation for candidate transients whose morphologies are not specified to the model in advance.

In simulated source-confused data, approximately 95\% or more of the MBHB samples exceeded $\mathcal{O}=0.9$ across SNRs of 30--50.  The same model recovered the dominant structures of instrumental glitches and cosmic-string bursts, neither of which was represented during training.  Across these two cross-morphology classes, the fractions exceeding $\mathcal{O}=0.8$ were typically above 75\% over the same SNR range.  The neighboring-subsample analysis further showed that the network mapped complementary noisy views to consistent signal-coherent representations.

Once auxiliary data or payload-monitoring data independently identify a candidate transient as instrumental, the framework can estimate and subtract its waveform without an anomaly-specific template.  Such conditional subtraction assumes that the anomaly contains no astrophysical or cosmological signal.  In the simulated 30-day stream, subtraction suppressed 69.8\% of the injected-anomaly power over the full analyzed frequency range and 87.1\% within the glitch-dominated band.

The present validation is limited to simulated data and two cross-morphology transient classes.  Data gaps and nonstationary noise conditions have not yet been evaluated.  Extending the validation to broader transient populations and more realistic observational artifacts is therefore an important next step.  Within these bounds, the results support self-supervised extraction for initial waveform estimation, followed by transient characterization or conditional subtraction of independently identified instrumental anomalies.

%% file: sections/07_acknowledgments.tex
\begin{acknowledgments}

This work is supported by the National Key Research and Development Program of China under Grant Nos. 2025YFE0217300, 2021YFC2201901, and 2021YFC2201903.

\end{acknowledgments}